\documentclass[sigconf]{acmart}

\usepackage{pifont}
\usepackage{tabularx}
\usepackage{enumerate,enumitem}
\usepackage{hyperref}
\usepackage{url}
\usepackage{array}
\usepackage{xcolor}

\begin{document}

\title{Non-Binary LDPC Arithmetic Error Correction For Processing-in-Memory}

\author{Daijing Shi$^{\triangle}$,
	Yihang Zhu$^{\triangle}$,
	Anjunyi Fan$^{\triangle}$,
	Yaoyu Tao$^{\triangle *}$,
	Yuchao Yang$^{\triangle\square*}$,
	and Bonan Yan$^{\triangle *}$
	}
\email{{taoyaoyutyy,yuchaoyang,bonanyan}@pku.edu.cn}
\affiliation{
\institution{$^{\triangle}$Peking University  $^{\square}$Chinese Institute for Brain Research (CIBR)}
\city{Beijing}
\country{China}
}

\settopmatter{printacmref=false} 
\renewcommand\footnotetextcopyrightpermission[1]{} 
\pagestyle{plain} 

\renewcommand{\shortauthors}{Shi et al.}

\begin{abstract}
Processing-in-memory (PIM) based on emerging devices such as memristors is more vulnerable to noise than traditional memories, due to the physical non-idealities and complex operations in analog domains. To ensure high reliability, efficient error-correcting code (ECC) is highly desired. However, state-of-the-art ECC schemes for PIM suffer drawbacks including dataflow interruptions, low code rates, and limited error correction patterns. In this work, we propose non-binary low-density parity-check (NB-LDPC) error correction running over the Galois field. Such NB-LDPC scheme with a long word length of 1024 bits can correct up to 8-bit errors with a code rate over 88\%. Nonbinary GF operations can support both memory mode and PIM mode even with multi-level memory cells. We fabricate a 40nm prototype PIM chip equipped with our proposed NB-LDPC scheme for validation purposes. Experiments show that PIM with NB-LDPC error correction demonstrates up to 59.65$\times$ bit error rate (BER) improvement over the original PIM without such error correction. The test chip delivers 2.978$\times$ power efficiency enhancement over prior works.
\end{abstract}

\maketitle
\section{Introduction}\label{sec:intro}

Processing-in-memory (PIM) technologies have by far reshaped the chip architecture design by enabling highly efficient in-/near-memory computations, especially for accelerating multiply-and-accumulation (MAC) kernels dominant in deep neural networks (DNNs)~\cite{zheng2024improving}. 
A handful of PIM designs have been developed based on different kinds of volatile memories such as static random-access memory (SRAM)~\cite{fu2023probabilistic}, dynamic random-access memory (DRAM)~\cite{kao2024ultra}. Recently, emerging nonvolatile memories such as memristors have shown great potential in further improving PIM's computing efficiency, including resistive random-access memory (RRAM)~\cite{wen2024fusion, yue2024scalable}, magnetic random-access memory (MRAM)~\cite{xie2024ed,zhang2024efficient} and phase-change memory (PCM)~\cite{le202364}. With PIM technologies, multiply-and-accumulation (MAC) operations can be executed inside the memory arrays by accumulating voltages~\cite{yoon202240,9731545}, charges~\cite{lee202328,aisy.201900068} or currents~\cite{huo2022computing,xie2024ed,8776485} along bitlines, etc. The overall computing efficiency is therefore boosted by avoiding repetitive data transfer between memories and external processing circuits.

\begin{figure}[t]
	\centering
	\includegraphics[width=1.0\columnwidth]{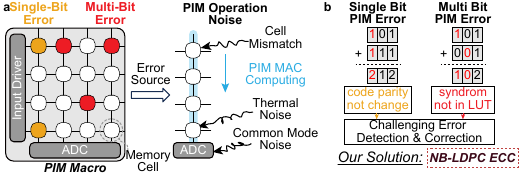}
	\caption{ (a) Noise degrades PIM computation accuracy. (b) Correction failures caused by limited supported error correction patterns.}
	\label{fig:1}
	\vspace{-2em}
\end{figure}

Despite the advantages of PIM, the accumulation along bitlines introduces unavoidable non-idealities regardless of the underlying memory devices. The noise introduced by them is injected probabilistically into data stored in the memory cells and the output computing results~\cite{fan2012analysis,yue2024scalable,he2019noise, zhang2019design,yang2020retransformer,yue2025physical}, such as thermal noise and flickering noise.
In particular, emerging memories may introduce higher device variations and the resulting hard errors further deteriorate the bit error rates (BER) for processing in or near memories, which are easy to induce multi-bit errors (Fig.~\ref{fig:1}(a)).

To address this unavoidable challenge, error-correcting codes (ECC) are the most commonly used method to attenuate noise corruptions, even though hardware-oriented retraining techniques have also been reported to mitigate the reliability problem of PIM towards DNN applications~\cite{chen2021pruning, shin2021fault, nguyen2021low, putra2021sparkxd, zhang2021efficient}. Integrating ECC with PIM computations requires the ECC to be compatible with PIM's intrinsic arithmetic operations. Existing methods can be roughly divided into two categories depending on whether the ECC needs to interrupt the PIM's computational dataflow. On the one hand, ECC that interrupts PIM computational dataflow usually detects errors and corrects the results by recursively reading the memory row by row~\cite{crafton2022improving}. On the other hand, existing ECCs without interrupting PIM computational dataflow need relatively large look-up tables to enable the decoding and error-correcting process~\cite{crafton2021cim, li202240nm}. Furthermore, existing methods are mainly based on short word lengths with relatively low code rates, which incur excessive area and power overheads due to the high parallelism required to support PIM processing. The error correction capability is by far up to 3-bit errors~\cite{crafton2022improving}. Due to the limited support of error correction patterns, existing methods also cannot support commonly used PIM design techniques, such as two-column differential weight mapping~\cite{xie2024ed, jing2022vsdca}.

In this work, we present a novel non-binary low-density parity-check (NB-LDPC) ECC scheme \cite{park2014fully,declercq2007decoding,2019automated, delima2024fault} optimized for PIM  designs to address the above challenges while achieving high code rate and energy efficiency. The main contributions of this work can be summarized as follows:
\begin{itemize}[leftmargin=*]
	\item We formulate a unified error detection and correction flow for PIM which can work in both conventional read/write mode and PIM mode. The proposed NB-LDPC is compatible with PIM MAC operations.
	\item We develop a novel NB-LDPC long-code-word ECC scheme, which is capable of correcting an arbitrary number of bit errors (depending on the iterative loops) without interrupting PIM computing dataflow.
	\item We tape out an RRAM-based PIM prototype chip monolithically integrated with the proposed NB-LDPC ECC module. To the best of our knowledge, this is the first attempt to realize PIM error correction with LDPC-type ECC. We also release the NB-LDPC decoder Verilog HDL implementation at \url{https://github.com/NoNameSubmission/NBLDPC_PIM}.
\end{itemize}

In addition to silicon-proven results for NB-LDPC PIM, we further investigate the design space of NB-LDPC with PIM circuitries. Measurements from the fabricated chip and simulations for design space explorations validate the effectiveness of the proposed NB-LDPC scheme. The results show that NB-LDPC ECC design improves the ECC power efficiency for PIM by 2.978$\times$ over existing designs in \cite{crafton2022improving,chang202240nm}. Furthermore, the NB-LDPC ECC is capable of achieving a code rate of more than 88\% with 1024-bit word length and correcting at most 8 error bits.

\section{Preliminaries}\label{sec:background}
\subsection{Processing-In-Memory Design}\label{sec:CIM}
Fig.~\ref{fig:1}(a) illustrates the fundamental PIM macro structure, including input drivers (address controllers), computing units, and analog-to-digital converters (ADCs). 
The PIM computation can be abstracted as:
\begin{equation}
	\mathbf{X}_{1\times n}\cdot\mathbf{W}_{n\times m}=\mathbf{Y}_{1\times m}
	\label{eq:0}
\end{equation}
where $\mathbf{X}$ is the input vector, $\mathbf{W}$ is the weight matrix, and $\mathbf{Y}$ is the output vector, $n$ and $m$ denotes the dimensions.
Before computation, the weight matrix \textbf{W} is binary-encoded and stored in memory arrays. The input vector of VMM operations is fed into PIM through input drivers. With a bit-serial scheme, each bit of the input elements is applied to one WL per clock cycle, controlling the on/off states of the memory cells. The computing results are then accumulated as analog signals on BLs. These signals are sampled and quantized by the ADCs for further operations such as shift-and-add~\cite{lee202328,chang202240nm} to generate the final digital outputs. During this computation process, the output analog signals are prone to noise, leakage current from SRAM cells, non-idealities of RRAM cells, or process variations, etc. (Fig.~\ref{fig:1}(a)), degrading the BER of the computing results. 

\subsection{Existing ECC designs for PIM}

Existing ECC designs can be categorized into two classes, one requiring dataflow interruption while the other one does not. For methods without dataflow interruption, they usually employ lookup tables (LUTs) for error detection and correction. Crafton, \textit{et al.}~\cite{crafton2021cim} propose a revised Hamming code that enables error correction of $\pm1$ bit in specific codeword. Li, \textit{et al.}~\cite{li202240nm} utilizes the modulo operation to compute the encoded word and the output error syndromes. However, large LUTs are difficult to scale up for longer word length or higher code rates, limiting the practical adoptions of these methods. On the other hand, ECC designs for PIM that require dataflow interruptions include ~\cite{crafton2022improving} that detects the error columns of the output words and corrects errors through repeatedly reading the information stored in the memory array. This method achieves good BER performance for RRAM parallel computing but adds extra timing overheads to PIM dataflow.

Fig.~\ref{fig:1}(b) shows the limited error patterns that can be corrected by the existing ECC designs, which usually consider the $\pm1$ error patterns. Thus, these kinds of ECC exhibit efficient error correction performance for a short word length, for example, 32 data bits. However, for a large-scale neural network, large memory capacity brings non-negligible space consumption for short-word-length ECC decoders and check bits. When differential operations are applied, the minimum length of the data bits is 2, covering ``-1'', ``0'', and ``1''. Since the existing methods consider only binary elements, they are difficult to process non-binary codewords. 

\section{Unified ECC For Memory \& PIM Modes}\label{Design}
\subsection{PIM Error Detection With NB-LDPC}\label{sec:Adapt}

\begin{figure}[t]
	\centering
	\includegraphics[width=1.0\columnwidth]{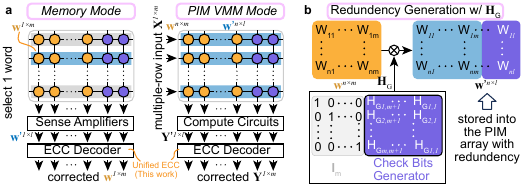}
	\caption{(a) Illustration of ECC operational flow for memories and PIM MAC. (b)The components of the generation matrix $\mathbf{H}_G$ and the hardware deployment of the encoded data bits and check bits (redundancy).}
	\label{fig:3}
\end{figure}

\begin{figure*}[t]
	\centering
	\includegraphics[width=2.0\columnwidth]{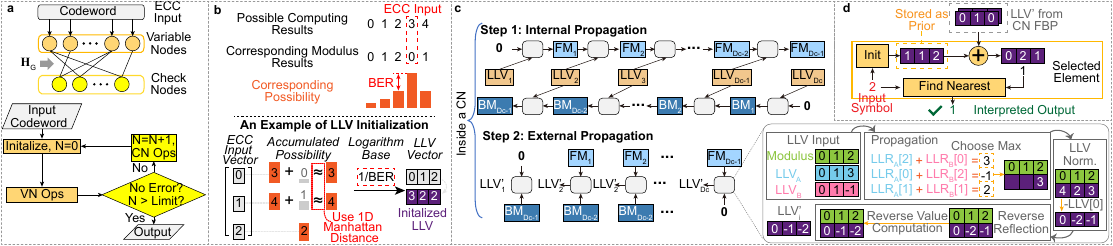}
	\caption{
		(a) The overall scheme of the NB-LDPC decoder.
		(b) LLV initialization process.
		(c) FBP algorithm.
		(d) Error Correction process in the VNs.
	}
	\label{fig:4}
\end{figure*}

We present an NB-LDPC coding design as a unified ECC scheme for both the memory mode and the PIM mode.
Fig.~\ref{fig:3}(a) shows the decoding flow of the proposed ECC. The NB-LDPC code can be described by a generation matrix $\textbf{H}_G$ and a check matrix $\textbf{H}_C$.
Their elements are all in a predefined Galois field of order $p$, i.e. \textit{GF}($p$).
The name of NB-LDPC originates from \textit{GF}($p$) when $p>2$, where the elements of $\mathbf{H}_G$ and $\mathbf{H}_C$ are not limited to binary elements.
$\textbf{H}_G$ and $\textbf{H}_C$ can be generated by the LDPC construction methods, e.g. PEG~\cite{venkiah2008design} and PCEG~\cite{han2020minimum}. 
$\mathbf{H}_G$ is an $m\times m$ identity matrix concatenated with an $m\times l$ matrix for the check bits generation (Fig.~\ref{fig:3}(b)). $\mathbf{H}_C$ is thereby generated to satisfy:
\begin{equation}
	\mathbf{H}_{G}\cdot\mathbf{H}_{C}^\intercal = \mathbf{0}
	\label{eq:1}
\end{equation}
where $\mathbf{0}$ is the zero matrix. Subsequently, we formulate the workflow of NB-LDPC for PIM designs.

\textbf{\textit{Memory Mode Error Detection}:} When sensing $m$ original data bits and forming a vector\footnote{Here the data bits can be binary bits or non-binary numbers in \textit{GF}(p).} $\textbf{w}^{1\times m}$ (the superscript denotes its dimension), the actual data that are stored in the PIM array is ${\mathbf{w'}}^{1\times l}=\mathbf{w}^{1\times m}\cdot\mathbf{H}_G^{m\times l}$ (Fig.~\ref{fig:3}(b)).
$l$ is the vector length of data bits plus check bits, thus $l>m$.
Note that the code rate is defined as $\frac{m}{l}\le 1$ where $m$ is the number of data bits.
To detect errors, a checking operation is defined as multiplying the codeword under test by the check matrix, i.e. $\left(\mathbf{w'}\cdot\mathbf{H}_C^\intercal\right)$. There are no errors detected if and only if:
\begin{equation}
	{\mathbf{w'}}^{1\times l}\cdot\left(\mathbf{H}_C^{m\times l}\right)^\intercal=\mathbf{0}
	\label{eq:2}
\end{equation}

\textbf{\textit{PIM Mode Error Detection}:} In PIM MAC operations with the NB-LDPC code, both MAC for data bits and check bits are performed. Eq.~\ref{eq:0} is revised to involve the check bits:
\begin{equation}
	\mathbf{X}^{1\times n}\cdot\mathbf{W'}^{n\times l}=\mathbf{Y}'^{1\times l}
	\label{eq:3}
\end{equation}
Thanks to the linearity, $\mathbf{Y}'^{1\times l}$ still satisfies the check matrix operation that yields a zero matrix \textbf{if and only if} there is no error in the PIM MAC results (Eq.~\ref{eq:4}). If the check matrix operation leads to non-zero results, error bits are detected.
\begin{equation}
	\mathbf{Y}'^{1\times l}\cdot(\mathbf{H}_C^{m\times l})^\intercal=\mathbf{X}^{1\times n}\cdot\mathbf{W'}^{n\times l}\cdot(\mathbf{H}_C^{m\times l})^\intercal=\mathbf{0}
	\label{eq:4}
\end{equation}

\subsection{PIM Error Correction With NB-LDPC}\label{sec:IntDecode}
If error bits are detected, the NB-LDPC decoder will then try to correct them.
Fig.~\ref{fig:4}(a) depicts the error correction process of the proposed NB-LDPC scheme.
It consists of variable nodes (VNs), check nodes (CNs), and the interconnections between them that are determined by the check matrix $\mathbf{H}_C$. 
VNs process the received codewords from the PIM array (i.e. $\mathbf{Y}'^{1\times l}$) and determine the final corrected output. CNs calculate the error syndrome and execute the iterative operations to progressively correct and decode the words. The decoding process of the proposed NB-LDPC consists of three steps:

\subsubsection{Logarithmic Likelihood Value Initialization}\label{sec:Init}

When an encoded codeword $\mathbf{Y}'^{1\times l}$ is fed into the NB-LDPC ECC decoder, for each bit, a group of logarithmic likelihood values (LLVs) that stands for the confidence levels of input being each element $k\in$\textit{GF}($p$) is computed inside each VN. The LLV of element $k$ is defined as the logarithmic value of the posterior probability that the correct computing result is $k$.
In this design, we simplify LLV computation as the one-dimensional Manhattan distance from the element $k$ to the original input code bit (Fig.~\ref{fig:4}(b)).
Based on our design for the prototype chip (Sec.~\ref{sec:hi}), this tradeoff incurs slight BER degradation from $3.57\times10^{-6}$ to $4.98\times10^{-6}$ but saves 21.65\% area and 23.69\% power consumption compared with design using full precision computation for LLV computations.

\subsubsection{Forward-Backward Propagation}\label{sec:FBP}

The LLVs generated in Sec.~\ref{sec:Init} are stored in VNs as the prior LLVs to start the iterative decoding process. To start the iteration, the prior LLVs in VNs are treated as the temporal LLVs of the 0$^{\mathrm{th}}$ iteration.
The temporal LLVs are then sent to CNs according to $\mathbf{H}_{C}$ acting as the adjacency matrix.
During the sending process, the LLVs of the VNs are sorted by the element of the traveling paths.
Specifically, the LLV corresponding to element $k \in $\textit{GF}($p$) from the i$^\mathrm{th}$ VN will be stored in the $(k\cdot H_{C_{ij}})^\mathrm{th}$ LLV of the j$^\mathrm{th}$ CN (Eq.~\ref{eq:6}) using the following index function:
\begin{equation}
	\mathbf{LLV}_{j}[k]=\mathbf{LLV}_{i}[k\cdot H_{C_{ij}}]
	\label{eq:6}
\end{equation}
$\mathbf{LLV}$ is the iterated vector of LLVs, $H_{C_{ij}}$ is the ($i$,$j$)$^{\mathrm{th}}$ element of $\mathbf{H}_C$.
In CNs, the received LLVs will be used to detect errors and generate information for temporal LLVs update in VNs.
An algorithm called Forward-Backward Propagation (FBP) is applied in the CNs~\cite{davey1998low,wymeersch2004log}.
FBP is to generate a new LLV group for each VN$_i$ connecting to the same CN, which can tell the specific VN what the other VNs think the correct code bit of it should be.
FBP is done in two steps:

\textbf{Step 1-Internal Propagation:}
In each CN, the LLVs from different VNs are transmitted in two directions to generate forward messages (FM) and backward messages (BM).
The propagation process shown in Fig.~\ref{fig:4}(c) takes two groups of LLVs as input.
In this step of the $i$-th propagation, the input will be \textbf{FM}$_{i-1}$ or \textbf{BM}$_{i-1}$ and \textbf{LLV}$_{i}$ or \textbf{LLV}$_{D_{C}-i+1}$ depending on the directions, respectively.
These two groups of LLVs are first ``added'' in the logarithmic domain to form a new group, corresponding to the multiplication of probabilities.
The addition result of the LLVs for each element $k$ is determined by the maximum LLV summation of the possible choices (Eq.~\ref{eq:8}).
\begin{equation}
	\mathbf{LLV}_{o}[k]=\max\{j\in \textit{GF}(p)\mid \mathbf{LLV}_{A}[k-j]+\mathbf{LLV}_{B}[j]\}
	\label{eq:8}
\end{equation}
where $\mathbf{LLV}_{A}$, $\mathbf{LLV}_{B}$ are the input of the propagation module, and $\mathbf{LLV}_{o}$ is the output of it. 
The generated LLVs of each element are then normalized to prevent the accumulation effect by subtracting the LLV of element ``0'' from all of the values in the group.
After that, the LLVs are reflected to its reverse element in \textit{GF}($p$), which will be the LLV group of the next FM or BM.
Completing the generation of all FMs and BMs starts the following \emph{Step 2}.

\textbf{Step 2-External Propagation:}
As shown in Fig.~\ref{fig:4}(c), external propagation is based on the previously generated FMs and BMs.
The i$^\mathrm{th}$ LLV$'$ in the CN is propagated from the FM$_{i-1}$ and BM$_{D_C-i}$ for temporal LLV updating in VN$_i$.
As mentioned in \emph{Step 1}, in this operation, the i$^\mathrm{th}$ LLV$'$ is generated by all the VNs connected to the CN except from i$^\mathrm{th}$ VN.
Thus, the final $\mathbf{LLV}'_i$ transmitted back to VN$_i$ eliminates the information from the specific node, preventing the repeated strengthening of the prior LLVs.
Transmission of the LLV$'$s from CNs to VNs is according to Eq.~\ref{eq:6}, which stands for the end of operations in CNs in this iteration.

\subsubsection{Accumulative Error Correction}\label{sec:Correct}

After receiving the LLV$'$s from corresponding CNs, the VNs will start updating the temporal LLV groups determining the final output of the decoder in this iteration.
The prior LLVs generated by the initialization process are added to the LLV$'$s sent back to the VNs by a normal operation instead of that in Sec.~\ref{sec:FBP}.
The element corresponding to the largest LLV in this updated group stands for the final corrected result in \textit{GF}($p$) (Fig.~\ref{fig:4}(d)).

After decoding the codeword in \textit{GF}($p$), the corrected computing results of the PIM units are subsequently interpreted.
The interpreted word is the one that has the shortest 1D Manhattan distance to the determined code bit in this digit.

\subsection{NB-LDPC Compatibility With Various PIM Schemes}
We develop the proposed PIM NB-LDPC ECC scheme in consideration of supporting multi-level cell memory and multi-bit integer-based PIM MAC arithmetic because the proposed NB-LDPC arithmetic code is built on non-binary \textit{GF}($p$) (where $p$ is a prime number). As a specific case, the widely-used differential weight mapping technique~\cite{xie2024ed, jing2022vsdca} for analog-computing PIM can be regarded as a ternary element utilizing the characteristics of modulo operations on negative numbers. Compared with conventional ECC for PIM that only supports binary elements, the proposed design paves the way to implement error correction for general PIM designs. 

\section{Circuit Architecture of NB-LDPC}\label{sec:circuit}
\begin{figure}[t]
	\centering
	\includegraphics[width=1.0\columnwidth]{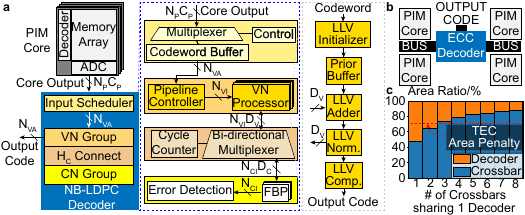}
	\caption{(a) Digital circuit architecture of the proposed NB-LDPC decoder. (b) The PIM memory architectures of multiple PIM cores sharing one ECC decoder. (c) NB-LDPC area overhead in scaled-out PIM memory architectures.}
	\label{fig:5}
\end{figure}

Fig.~\ref{fig:5}(a) shows the circuit architecture of the proposed NB-LDPC arithmetic code. Porting from PIM cores, the proposed NB-LDPC mainly includes an input scheduler, VN and CN processing units, and multiplexers determined by $\mathbf{H}_C$. The design parameters are presented in Table~\ref{tab:0}. 
To conserve chip space, hardware implementations frequently use only partial CNs and VNs, reusing them across time steps.($N_{VI}$, $N_{VA}$) and  ($N_{CI}$, $N_{CA}$) represent the number of VNs and CNs realized in hardware and defined in the NB-LDPC algorithm, respectively. $D_V$ and $D_C$ denote the sparse connectivity of each node in VNs and CNs to each other defined in $\mathbf{H}_C$.

\begin{table}[t]
	\caption{Parameter List for PIM+NB-LDPC}
	\begin{center}
		\begin{tabularx}{1\columnwidth} { 
				>{\centering\hsize=.13\hsize\linewidth=\hsize}X 
				| >{\hsize=.60\hsize\linewidth=\hsize}X 
				| >{\hsize=.14\hsize\linewidth=\hsize}X}
			\hline
			\hline
			Parameter & Meaning & Typical Values\\
			\hline
			$N_P$ & Number of PIM cores sharing 1 ECC decoder & 1$\sim$64\\
			\hline
			$C_P$ & Column parallelism$^\dagger$ of each PIM core & 1$\sim$1024\\
			\hline
			$N_{VI}$ & Number of VNs for iterative decoding & 32$\sim$1024\\
			\hline
			$N_{VA}$ & Number of VNs designed in the algorithm & 32$\sim$1024\\
			\hline
			$N_{CI}$ & Number of CNs for iterative decoding & 1$\sim$16\\
			\hline
			$N_{CA}$ & Number of CNs designed in the algorithm & 4$\sim$512\\
			\hline
			$D_{V}$ & VNs connection degrees & 2$\sim$4\\
			\hline
			$D_{C}$ & CNs connection degrees & 6$\sim$18\\
			\hline
			\hline
			\multicolumn{3}{l}{$^\dagger$Column parallelism is defined as the number of ADCs in a PIM macro.}
		\end{tabularx}
	\end{center}
	\label{tab:0}
\end{table}

The principle above for NB-LDPC error correction forms a datapath as follows: 
\ding{182} The generated codewords from the $N_P$ PIM cores (i.e. PIM MAC computing results) with column parallelism $C_P$ are firstly fed into the input schedulers. The input scheduler buffers and generates the codewords in multiple cycles in a predefined order. They are then sent to the VNs for LLV initialization where the initialized prior LLVs are stored (Sec.~\ref{sec:Init}).  
\ding{183} Transmission between the VNs and CNs is defined by $\mathbf{H}_C$ fixed as hardware connections of multiplexers. According to the ratio of $N_{P}C_{P}/N_{VI}$ and $N_{CI}/N_{CA}$, the LLV initialization cycles and the required cycles of the iterative decoding can be determined, respectively.
\ding{184} The LLVs transmitted to CNs then go through the FBP algorithm. This propagation process is realized by a dedicated computing module embedded inside the CNs to compute both FM or BM and generate LLV$'$s (Sec.~\ref{sec:FBP}). Note that error detection is naturally executed in CNs in that the summation of the codewords is naturally completed during the propagation process. 
\ding{185} At the end of the propagation, the FM or BM is compared with the last or the first LLV group sent by the VNs. Once the maximum element of the 2 groups is the same, elements in this check node pass the error detection. If all the check nodes pass the error detection process, the decoding iteration is over. Otherwise, the LLV$'$s should be sent back to the VNs and start the next iteration. The final codeword generation is carried out in the VNs (Sec.~\ref{sec:Correct}).

\section{Hardware Prototype of PIM With NB-LDPC}\label{sec:hi}

To validate the effectiveness and efficiency of the proposed NB-LDPC, we design and fabricate a prototype chip in a commercial 40nm technology node. It monolithically integrates an RRAM PIM macro with 2.5-bit flash ADCs and a proposed NB-LDPC ECC module. A 256$\times$320 binary 1-RRAM-1-transistor (1T1R) array, namely $N_P=1$ is embedded as the PIM core of the design, with $C_{P}=10$ using Flash ADCs to convert and quantize the analog computing results. For the NB-LDPC realization, we use \textit{GF}(3), resulting in each check bit length of 2 bits. The on-chip NB-LDPC decoder contains 288 VNs ($N_{VI}=288$) and 1 CN ($N_{CI}=1$) to compromise the throughput of the PIM core. Since $N_{VI}=N_{VA}$, LLV initialization and arrangement are controlled by digital logic. The operations of the CNs and transmission between VNs and CNs are then controlled by a finite state machine (FSM) and multiplexers. Besides, the design is also facilitated with a debug port for debug codeword input. We release the proposed PIM NB-LDPC ECC implementation source code in Verilog HDL in \url{https://github.com/NoNameSubmission/NBLDPC_PIM}.

\section{Evaluation}\label{sec:Evaluate}

\subsection{Experiment \& Simulation Setup}\label{sec:setup}

\begin{figure}[t]
	\centering
	\includegraphics[width=1.0\columnwidth]{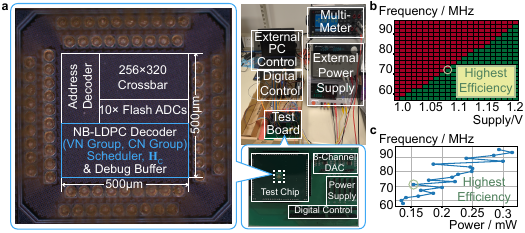}
	\caption{(a) Our prototype chip: RRAM PIM \& monolithically integrated NB-LDPC decoder @ 40nm technology node and chip test environments. (b) \textbf{Measured} Shmoo plot of the prototype chip. (c) \textbf{Measured} average power consumption of NB-LDPC module solely (excluding RRAM \& ADCs).}
	\label{fig:7}
\end{figure}

Evaluation of the proposed PIM NB-LDPC includes the measurement of the prototype chip and the cycle-accurate simulation for design exploration.
\ding{182} \textbf{Chip Measurement:}
as shown in Fig.~\ref{fig:7}(a), the presented testing chip is embedded into a test board for the experiment.
An FPGA DE0-CV development board is applied as the signal controller of the test chip.
The debug port of the chip is connected to the PC using an I$^2$C /SPI host adapter working in the I$^2$C mode for data transportation.
To measure the power consumption of the ECC decoder, we first measure the power consumption of the digital circuits working in the decoding process.
Then, the power consumption of the data transportation port and the digital debug buffer is measured under the same condition.
The power consumption of the proposed ECC design is obtained by subtracting the power consumption of debug circuits from that of the total chip.
\ding{183} \textbf{Design Space Exploration Simulation:}
In addition to the prototype chip as one possible implementation, this work also explores the large design space. To explore the influence of the design parameters in Table~\ref{tab:0}, we synthesize the decoder circuits with different parameters and perform Verilog-based cycle-accurate post-synthesis simulation.
Further, we simulate the influence of the NB-LDPC algorithmic parameters on the BER improvement under various implementations.
This algorithm-level simulation is done using Numpy in Python.
The construction of the check matrix and generator matrix is based on existing coding algorithmic works~\cite{han2020minimum,venkiah2008design}.
The specific value of these non-zero elements of $\mathbf{H}_C$ are randomly picked from the non-zero value in \textit{GF}($p$) where the NB-LDPC ECC for PIM is built.

The BER performance of the NB-LDPC arithmetic code is measured and simulated for different cases.
The hardware-implemented results of the design under different raw BER are measured utilizing the debug port mentioned in Sec.~\ref{sec:hi}.
Other experiments evaluating the influence of code rates and word lengths are demonstrated through simulations.
To show the performance of the proposed ECC design in large-scale neural network applications, a simulation of ResNet-34 on ImageNet dataset~\cite{he2016deep} is carried out with the proposed ECC.
With the help of ~\cite{zhou2016dorefa, liu2020reactnet}, we quantize the DNN weights and activating values of the first and the last layer in the model to 8-bit to fit the PIM scheme, while the other layers are quantized to ternary weight and binary inputs.

\begin{table}[t]
	\caption{Comparison of PIM ECC Designs}
	\centering
	\begin{tabularx}{1\columnwidth}{ 
			>{\centering\hsize=.28\hsize\linewidth=\hsize}X 
			| >{\hsize=.16\hsize\linewidth=\hsize}X 
			| >{\hsize=.08\hsize\linewidth=\hsize}X
			| >{\hsize=.08\hsize\linewidth=\hsize}X
			| >{\hsize=.20\hsize\linewidth=\hsize}X
			| >{\hsize=.16\hsize\linewidth=\hsize}X
		}
		\hline
		\hline
		Work & Row Parallelism & MWL$^\dagger$ (bits) & MTE$^\ddagger$ (bits) & ECC Efficiency (Mbps/W)$^\triangle$ & Efficiency Improvement\\
		\hline
		\textbf{This work} & \textbf{Arbitrary} & \textbf{256} & \textbf{5} & \textbf{1152.00}$^{*}$ & 2.978$\times$\\
		\hline
		DAC'22$^{\text{~\cite{chang202240nm,crafton2022improving}}}$ & 8 & 32 & 3 & 386.82 & 1$\times$\\
		\hline
		ASSCC'21$^{\text{~\cite{crafton2021cim}}}$ & 4 & 32 & 1 & 35.92 & 0.093$\times$\\
		\hline
		ESSCIRC'22$^{\text{~\cite{li202240nm}}}$& 7 & 25 & 1 & 88.47 & 0.229$\times$\\
		\hline
		\hline
		\multicolumn{6}{l}{$^{\dagger}$ MWL means Maximum Word Length}\\
		\multicolumn{6}{l}{$^{\ddagger}$ MTE means Maximum Tolerable Errors in the output word}\\
		\multicolumn{6}{l}{$^{\triangle}$ ECC power efficiency is computed through $\frac{\text{Corrected Bit Rate}}{\text{Power}}$}\\
		\multicolumn{6}{l}{$^{*}$ Power for comparison is measured under row parallelism of 4}\\
	\end{tabularx}
	\label{tab:1}
	
\end{table}

\subsection{Measured Prototype Chip ECC Hardware Performance}\label{sec:measure}
As shown in Fig.~\ref{fig:7}(b), the performance of the proposed design at a word length of 256 bits and 80\% code rate is measured using the prototype chip mentioned in Sec.~\ref{sec:hi}.
The working states are measured from 1V to 1.2V, with the ECC decoder continuously working.
The main clock frequency of the prototype chip varies from 58MHz to 95MHz, validating the feasibility of the proposed NB-LDPC ECC for PIM.
Fig.~\ref{fig:7}(c) shows the mean NB-LDPC power consumption (x-axis, average from 1000 times measurement) under different clock frequencies (y-axis).
The data points in Fig.~\ref{fig:7}(b) correspond to the cases of the highest clock frequency under different supply voltages in the Shmoo plot.
With increasing clock frequency, the power consumption of the ECC decoder shows a growing trend.
The jitter of the measured curve is caused by the measurement method mentioned in Section~\ref{sec:setup}.
Since the power consumption of the debugging module is much larger than the decoder in our prototype chip, the measured results are influenced by the precision of the measurement.

``ECC power efficiency'' is defined as the corrected data bits throughput per unit power consumption~\cite{ferraz2021survey}.
Of our chip, the best ECC power efficiency point is measured at a 1.07V supply and 71MHz working frequency, resulting in 1152.00Mbps/W, as shown in Table~\ref{tab:1}.
For comparison, we provide the simulation or measured results reported by other works. The proposed design successfully improves the ECC efficiency by at most 2.978$\times$ and 2.382$\times$ on average of all cases compared with the best existing ECC for PIM~\cite{crafton2022improving}.
In addition, Table~\ref{tab:1} presents the advantage of the design that it has no requirements for row parallelism (i.e., how many rows are simultaneously turned on for PIM VMM) and supports a longer word length with multi-bit error-correcting ability.

\begin{figure}[t]
	\centering
	\includegraphics[width=1.0\columnwidth]{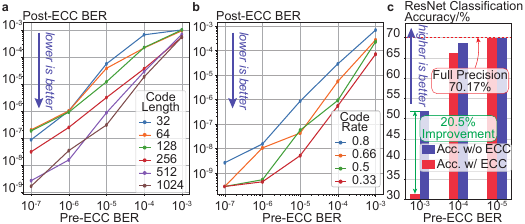}
	\caption{
		(a) NB-LDPC code performance with the same code rate but different word lengths.
		(b) NB-LDPC code performance with different code rates but the same word length.
		(c) DNN (ResNet-34) accuracy benchmark on PIM with or without the ECC, compared with the full precision results.
	}
	\label{fig:9}
\end{figure}

As shown in Fig.~\ref{fig:5}(c), sharing the ECC decoder among multiple cores can release the area cost to a large degree.
With 6 PIM macros sharing one decoder, the area penalty of the decoder will fall to lower than 25\%, which is lower than the reported results in ~\cite{crafton2022improving}.

\subsection{Simulated NB-LDPC Arithmetic Code Performance}\label{sec:sim}

Fig.~\ref{fig:9} verifies the error-correcting capability of NB-LDPC under different BERs.
The x-axis and y-axis are the pre- and post-ECC BERs, respectively.
Fig.~\ref{fig:9}(a) shows the impact of different NB-LDPC word lengths on the coding performance.
Besides the prototype chip (256-bit codeword), we generate multiple NB-LDPC codes with word lengths varying from 32 to 1024 (i.e. different $\mathbf{H}_G$ \& $\mathbf{H}_C$ pairs) with a fixed 80\% code rate for comparison.
Generally, in our design, shorter codes yield worse error correction performance than the longer ones under the same input BER.
Longer codewords reduce the likelihood of encountering specific dilemmas that can lead to error correction failures, a problem also seen in conventional LDPC designs.
The best BER improvement is achieved with a word length of 1024 which improves the BER by 59.65$\times$ from $1\times10^{-5}$ to $1.676\times 10^{-7}$.
Theoretically, the upper limit of the code rate can be up to 88\% with an information word length of 1024, but it has a trade-off with the BER performance.
Fig.~\ref{fig:9}(b) shows the influence of different NB-LDPC code rates on the coding performance.
The word length is fixed as 512 and the code rate varies from 0.33 to 0.8.
The results reveal that lower code rates inherently increase the redundancy of the code, making error correction more effective at the cost of more decoding overhead. 

To evaluate ECC benefits for an end-to-end PIM-based DNN computation, we simulate ResNet-34 computation with PIM executing its MAC.
Fig.~\ref{fig:9}(c) shows the simulated BER improvement using realistic PIM with and without NB-LDPC ECC.
The fault model is simplified and abstracted to a fixed probability of bit flip rate (x-axis) during computation ranging from $10^{-3}$ to $10^{-5}$.
The error might happen in both weights and activations.
The y-axis denotes the final ResNet-34 classification accuracy over the test dataset.
Results reveal that NB-LDPC brings significant improvement when BER is worse than $10^{-4}$.
With a BER at $1\times10^{-3}$, the performance of the neural network decreases by an absolute value of 38.54\%.
NB-LDPC (1024-bit codeword, 80\% code rate) significantly reduces the BER to $4.14\times10^{-4}$ and recovers the DNN classification accuracy by 20.5\%.

\subsection{Design Space Exploration for NB-LDPC Hardware}
\begin{figure}[t]
	\centering
	\includegraphics[width=1.0\columnwidth]{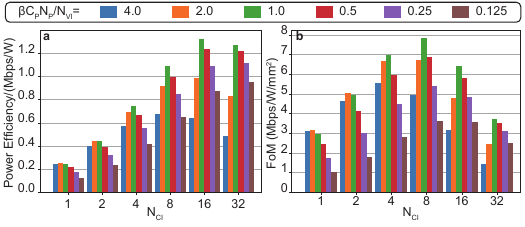}
	\caption{Simulated design space exploration for different parameters: (a) Power efficiency of NB-LDPC decoder, (b) figure of merit of NB-LDPC decoder.}
	\label{fig:10}
\end{figure}

As mentioned in Sec.~\ref{sec:circuit}, $N_{P}C_{P}/N_{VI}$ (implying the processing speed of the VN group) and $N_{CI}/N_{CA}$ (indicating the processing speed of CN group) are the two most important factors influencing the decode speed and power consumption.
Here we take the classical value of $N_{P}=4, C_{P}=10$ for design space exploration.
The code design is fixed and the same as in our silicon-proven prototype described in Sec.~\ref{sec:hi}.
Fig.~\ref{fig:10}(a) presents the influence of $N_{P}C_{P}/N_{VI}$ and $N_{CI}/N_{CA}$ on the ECC power efficiency.
Using the pre-determined parameters in the software algorithm, these two parameters can be changed into $\beta N_{P}C_{P}/N_{VI}$ and $N_{CI}$, where $\beta$ is the ratio of the VNs and the specific partial sums that need to be computed to generate an encoded codeword. In our simulated experiments, $\beta =(N_{VA}+N_{CA}) / (N_{VA}+2N_{CA})$ since each check bit requires 2 bits.
As can be seen from each group of the bar with the same $N_{CI}$, the highest power efficiency of the decoder always occurs when $\beta N_{P}C_{P}/N_{VI}=1$.
This is logical because no hardware will be suspended during the initialization process and thus the highest power efficiency should be achieved. 
Similarly, the highest power efficiency can be achieved when $N_{CI}=16$, namely $N_{CI}/N_{CA}=1$ in the simulation condition.

Further, the influences of these parameters on the overall figure of merit (FoM) of this design are explored.
The selected FoM is the ratio of ECC power efficiency over the NB-LDPC circuits area, which attempts to find the Pareto optimality of ECC power efficiency and the area overhead considering the BER of the whole system will not be affected by these hardware-implemented parameters.
Fig.~\ref{fig:10}(b) presents the simulated FoM results (y-axis) versus different $N_{CI}$ (x-axis).
As can be seen, there lies a sweet point to reach a co-optimal point for efficiency and design (area) overhead to maintain a high error correction capability.
The highest FoM value is achieved when $N_{CI}=8$, $\beta N_{P}C_{P}/N_{VI}=1$.
This is because of the higher cost of the CN processing unit than VN.
The former is 61.83$\times$ larger than the latter according to the synthesis results.

\section{Conclusion}\label{sec:conclude}
In this paper, we develop a novel NB-LDPC ECC for PIM that aims at high power efficiency and high code rate. With silicon-proven architecture and circuit designs, the power efficiency of the proposed NB-LDPC decoder is measured to be 2.978$\times$ better than the existing ECC designs~\cite{crafton2022improving}. Exploration and analysis of the proposed ECC design shows that, with the ability to detect and correct multiple errors in a word, a 1024-bit codeword can achieve a code rate of more than 88\% according to existing mathematical analysis. Simulation results show the proposed design can achieve 59.65$\times$ BER improvement compared with raw results with a word length of 1024 and 80\% code rate.

\bibliographystyle{ACM-Reference-Format}
\bibliography{refs.bib}


\begin{thebibliography}{41}


\ifx \showCODEN    \undefined \def \showCODEN     #1{\unskip}     \fi
\ifx \showISBNx    \undefined \def \showISBNx     #1{\unskip}     \fi
\ifx \showISBNxiii \undefined \def \showISBNxiii  #1{\unskip}     \fi
\ifx \showISSN     \undefined \def \showISSN      #1{\unskip}     \fi
\ifx \showLCCN     \undefined \def \showLCCN      #1{\unskip}     \fi
\ifx \shownote     \undefined \def \shownote      #1{#1}          \fi
\ifx \showarticletitle \undefined \def \showarticletitle #1{#1}   \fi
\ifx \showURL      \undefined \def \showURL       {\relax}        \fi
\providecommand\bibfield[2]{#2}
\providecommand\bibinfo[2]{#2}
\providecommand\natexlab[1]{#1}
\providecommand\showeprint[2][]{arXiv:#2}

\bibitem[Chang et~al\mbox{.}(2022)]%
        {chang202240nm}
\bibfield{author}{\bibinfo{person}{Muya Chang}, \bibinfo{person}{Samuel~D
  Spetalnick}, \bibinfo{person}{Brian Crafton}, \bibinfo{person}{Win-San Khwa},
  \bibinfo{person}{Yu-Der Chih}, \bibinfo{person}{Meng-Fan Chang}, {and}
  \bibinfo{person}{Arijit Raychowdhury}.} \bibinfo{year}{2022}\natexlab{}.
\newblock \showarticletitle{A 40nm 60.64{TOPS/W} {ECC}-Capable
  Compute-in-Memory/Digital {2.25MB/768KB} {RRAM/SRAM} System with Embedded
  {Cortex} {M3} Microprocessor for Edge Recommendation Systems}. In
  \bibinfo{booktitle}{\emph{IEEE International Solid-State Circuits Conference
  (ISSCC)}}, Vol.~\bibinfo{volume}{65}. IEEE, \bibinfo{pages}{1--3}.
\newblock


\bibitem[Chen and Chakrabarty(2021)]%
        {chen2021pruning}
\bibfield{author}{\bibinfo{person}{Ching-Yuan Chen} {and}
  \bibinfo{person}{Krishnendu Chakrabarty}.} \bibinfo{year}{2021}\natexlab{}.
\newblock \showarticletitle{Pruning of deep neural networks for fault-tolerant
  memristor-based accelerators}. In \bibinfo{booktitle}{\emph{ACM/IEEE Design
  Automation Conference (DAC)}}. IEEE, \bibinfo{pages}{889--894}.
\newblock


\bibitem[Crafton et~al\mbox{.}(2021)]%
        {crafton2021cim}
\bibfield{author}{\bibinfo{person}{Brian Crafton}, \bibinfo{person}{Samuel
  Spetalnick}, \bibinfo{person}{Jong-Hyeok Yoon}, \bibinfo{person}{Wei Wu},
  \bibinfo{person}{Carlos Tokunaga}, \bibinfo{person}{Vivek De}, {and}
  \bibinfo{person}{Arijit Raychowdhury}.} \bibinfo{year}{2021}\natexlab{}.
\newblock \showarticletitle{{CIM-SECDED}: A 40nm 64{Kb} Compute In-Memory
  {RRAM} Macro with {ECC} Enabling Reliable Operation}. In
  \bibinfo{booktitle}{\emph{IEEE Asian Solid-State Circuits Conference
  (A-SSCC)}}. IEEE, \bibinfo{pages}{1--3}.
\newblock


\bibitem[Crafton et~al\mbox{.}(2022)]%
        {crafton2022improving}
\bibfield{author}{\bibinfo{person}{Brian Crafton}, \bibinfo{person}{Zishen
  Wan}, \bibinfo{person}{Samuel Spetalnick}, \bibinfo{person}{Jong-Hyeok Yoon},
  \bibinfo{person}{Wei Wu}, \bibinfo{person}{Carlos Tokunaga},
  \bibinfo{person}{Vivek De}, {and} \bibinfo{person}{Arijit Raychowdhury}.}
  \bibinfo{year}{2022}\natexlab{}.
\newblock \showarticletitle{Improving compute in-memory {ECC} reliability with
  successive correction}. In \bibinfo{booktitle}{\emph{ACM/IEEE Design
  Automation Conference (DAC)}}. \bibinfo{pages}{745--750}.
\newblock


\bibitem[Davey and MacKay(1998)]%
        {davey1998low}
\bibfield{author}{\bibinfo{person}{MC Davey} {and} \bibinfo{person}{D MacKay}.}
  \bibinfo{year}{1998}\natexlab{}.
\newblock \showarticletitle{Low-density parity check codes over {GF(q)}}.
\newblock \bibinfo{journal}{\emph{IEEE Communications Letters}}
  \bibinfo{volume}{2}, \bibinfo{number}{6} (\bibinfo{year}{1998}),
  \bibinfo{pages}{165--167}.
\newblock


\bibitem[de~Lima et~al\mbox{.}(2024)]%
        {delima2024fault}
\bibfield{author}{\bibinfo{person}{Jo{\~a}o Paulo~Cardoso de Lima},
  \bibinfo{person}{Benjamin Franklin~Morris III}, \bibinfo{person}{Asif~Ali
  Khan}, \bibinfo{person}{Jeronimo Castrillon}, {and} \bibinfo{person}{Alex~K.
  Jones}.} \bibinfo{year}{2024}\natexlab{}.
\newblock \showarticletitle{Fault-Tolerant Masked Matrix Accumulation using
  Bulk Bitwise In-Memory Engines}.
\newblock \bibinfo{journal}{\emph{arXiv preprint arXiv:2409.10136}}
  (\bibinfo{year}{2024}).
\newblock


\bibitem[Declercq and Fossorier(2007)]%
        {declercq2007decoding}
\bibfield{author}{\bibinfo{person}{David Declercq} {and} \bibinfo{person}{Marc
  Fossorier}.} \bibinfo{year}{2007}\natexlab{}.
\newblock \showarticletitle{Decoding algorithms for nonbinary {LDPC} codes over
  {GF}$(q)$}.
\newblock \bibinfo{journal}{\emph{IEEE Transactions on Communications}}
  \bibinfo{volume}{55}, \bibinfo{number}{4} (\bibinfo{year}{2007}),
  \bibinfo{pages}{633--643}.
\newblock


\bibitem[Fan et~al\mbox{.}(2012)]%
        {fan2012analysis}
\bibfield{author}{\bibinfo{person}{Ming-Long Fan}, \bibinfo{person}{Vita Pi-Ho
  Hu}, \bibinfo{person}{Yin-Nien Chen}, \bibinfo{person}{Pin Su}, {and}
  \bibinfo{person}{Ching-Te Chuang}.} \bibinfo{year}{2012}\natexlab{}.
\newblock \showarticletitle{Analysis of Single-Trap-Induced Random Telegraph
  Noise on {FinFET} Devices, {6T} {SRAM} Cell, and Logic Circuits}.
\newblock \bibinfo{journal}{\emph{IEEE Transactions on Electron Devices (TED)}}
  \bibinfo{volume}{59}, \bibinfo{number}{8} (\bibinfo{year}{2012}),
  \bibinfo{pages}{2227--2234}.
\newblock


\bibitem[Ferraz et~al\mbox{.}(2021)]%
        {ferraz2021survey}
\bibfield{author}{\bibinfo{person}{Oscar Ferraz}, \bibinfo{person}{Srinivasan
  Subramaniyan}, \bibinfo{person}{Ramesh Chinthala}, \bibinfo{person}{Jo{\~a}o
  Andrade}, \bibinfo{person}{Joseph~R Cavallaro}, \bibinfo{person}{Soumitra~K
  Nandy}, \bibinfo{person}{Vitor Silva}, \bibinfo{person}{Xinmiao Zhang},
  \bibinfo{person}{Madhura Purnaprajna}, {and} \bibinfo{person}{Gabriel
  Falcao}.} \bibinfo{year}{2021}\natexlab{}.
\newblock \showarticletitle{A survey on high-throughput non-binary {LDPC}
  decoders: {ASIC}, {FPGA}, and {GPU} architectures}.
\newblock \bibinfo{journal}{\emph{IEEE Communications Surveys \& Tutorials}}
  \bibinfo{volume}{24}, \bibinfo{number}{1} (\bibinfo{year}{2021}),
  \bibinfo{pages}{524--556}.
\newblock


\bibitem[Fu et~al\mbox{.}(2023)]%
        {fu2023probabilistic}
\bibfield{author}{\bibinfo{person}{Yihan Fu}, \bibinfo{person}{Daijing Shi},
  \bibinfo{person}{Anjunyi Fan}, \bibinfo{person}{Wenshuo Yue},
  \bibinfo{person}{Yuchao Yang}, \bibinfo{person}{Ru Huang}, {and}
  \bibinfo{person}{Bonan Yan}.} \bibinfo{year}{2023}\natexlab{}.
\newblock \showarticletitle{Probabilistic Compute-in-Memory Design for
  Efficient Markov Chain Monte Carlo Sampling}.
\newblock \bibinfo{journal}{\emph{IEEE Transactions on Circuits and Systems I:
  Regular Papers (TCAS-I)}} (\bibinfo{year}{2023}).
\newblock


\bibitem[Han et~al\mbox{.}(2020)]%
        {han2020minimum}
\bibfield{author}{\bibinfo{person}{Changcai Han}, \bibinfo{person}{Hui Li},
  {and} \bibinfo{person}{Weigang Chen}.} \bibinfo{year}{2020}\natexlab{}.
\newblock \showarticletitle{Minimum Distance Optimization with Chord Edge
  Growth for High Girth Non-Binary {LDPC} Codes}.
\newblock \bibinfo{journal}{\emph{Electronics}} \bibinfo{volume}{9},
  \bibinfo{number}{12} (\bibinfo{year}{2020}), \bibinfo{pages}{2161}.
\newblock


\bibitem[He et~al\mbox{.}(2019)]%
        {he2019noise}
\bibfield{author}{\bibinfo{person}{Zhezhi He}, \bibinfo{person}{Jie Lin},
  \bibinfo{person}{Rickard Ewetz}, \bibinfo{person}{Jiann-Shiun Yuan}, {and}
  \bibinfo{person}{Deliang Fan}.} \bibinfo{year}{2019}\natexlab{}.
\newblock \showarticletitle{Noise injection adaption: End-to-end {ReRAM}
  crossbar non-ideal effect adaption for neural network mapping}. In
  \bibinfo{booktitle}{\emph{ACM/IEEE Design Automation Conference (DAC)}}.
  \bibinfo{pages}{1--6}.
\newblock


\bibitem[Henk et~al\mbox{.}(2004)]%
        {wymeersch2004log}
\bibfield{author}{\bibinfo{person}{Wymeersch Henk}, \bibinfo{person}{Steendam
  Heidi}, {and} \bibinfo{person}{Moeneclaey Marc}.}
  \bibinfo{year}{2004}\natexlab{}.
\newblock \showarticletitle{Log-domain decoding of {LDPC} codes over {GF(q)}}.
  In \bibinfo{booktitle}{\emph{IEEE International Conference on
  Communications}}, Vol.~\bibinfo{volume}{2}. IEEE, \bibinfo{pages}{772--776}.
\newblock


\bibitem[Huo et~al\mbox{.}(2022)]%
        {huo2022computing}
\bibfield{author}{\bibinfo{person}{Qiang Huo}, \bibinfo{person}{Yiming Yang},
  \bibinfo{person}{Yiming Wang}, \bibinfo{person}{Dengyun Lei},
  \bibinfo{person}{Xiangqu Fu}, \bibinfo{person}{Qirui Ren},
  \bibinfo{person}{Xiaoxin Xu}, \bibinfo{person}{Qing Luo},
  \bibinfo{person}{Guozhong Xing}, \bibinfo{person}{Chengying Chen},
  {et~al\mbox{.}}} \bibinfo{year}{2022}\natexlab{}.
\newblock \showarticletitle{A computing-in-memory macro based on
  three-dimensional resistive random-access memory}.
\newblock \bibinfo{journal}{\emph{Nature Electronics}} \bibinfo{volume}{5},
  \bibinfo{number}{7} (\bibinfo{year}{2022}), \bibinfo{pages}{469--477}.
\newblock


\bibitem[Kao et~al\mbox{.}(2024)]%
        {kao2024ultra}
\bibfield{author}{\bibinfo{person}{TC Kao}, \bibinfo{person}{MJ Huang},
  \bibinfo{person}{YR Liu}, \bibinfo{person}{YK Wang}, \bibinfo{person}{JC
  Guo}, {and} \bibinfo{person}{Steve~S Chung}.}
  \bibinfo{year}{2024}\natexlab{}.
\newblock \showarticletitle{An Ultra-Low Voltage {Auger}-Recombination Enhanced
  Hot Hole Injection Scheme in Implementing a 3 Bits per Cell e-{DRAM} {CIM}
  Macro for Inference Accelerator}. In \bibinfo{booktitle}{\emph{IEEE Symposium
  on VLSI Technology and Circuits (VLSI Symposia)}}. IEEE,
  \bibinfo{pages}{1--2}.
\newblock


\bibitem[Kim et~al\mbox{.}(2023)]%
        {he2016deep}
\bibfield{author}{\bibinfo{person}{Bokyung Kim}, \bibinfo{person}{Shiyu Li},
  {and} \bibinfo{person}{Hai Li}.} \bibinfo{year}{2023}\natexlab{}.
\newblock \showarticletitle{{INCA}: Input-stationary dataflow at
  outside-the-box thinking about deep learning accelerators}. In
  \bibinfo{booktitle}{\emph{IEEE International Symposium on High-Performance
  Computer Architecture (HPCA)}}. IEEE, \bibinfo{pages}{29--41}.
\newblock


\bibitem[Le~Gallo et~al\mbox{.}(2023)]%
        {le202364}
\bibfield{author}{\bibinfo{person}{Manuel Le~Gallo}, \bibinfo{person}{Riduan
  Khaddam-Aljameh}, \bibinfo{person}{Milos Stanisavljevic},
  \bibinfo{person}{Athanasios Vasilopoulos}, \bibinfo{person}{Benedikt
  Kersting}, \bibinfo{person}{Martino Dazzi}, \bibinfo{person}{Geethan
  Karunaratne}, \bibinfo{person}{Matthias Br{\"a}ndli},
  \bibinfo{person}{Abhairaj Singh}, \bibinfo{person}{Silvia~M Mueller},
  {et~al\mbox{.}}} \bibinfo{year}{2023}\natexlab{}.
\newblock \showarticletitle{A 64-core mixed-signal in-memory compute chip based
  on phase-change memory for deep neural network inference}.
\newblock \bibinfo{journal}{\emph{Nature Electronics}} \bibinfo{volume}{6},
  \bibinfo{number}{9} (\bibinfo{year}{2023}), \bibinfo{pages}{680--693}.
\newblock


\bibitem[Lee et~al\mbox{.}(2023)]%
        {lee202328}
\bibfield{author}{\bibinfo{person}{Kyeongho Lee}, \bibinfo{person}{Joonhyung
  Kim}, {and} \bibinfo{person}{Jongsun Park}.} \bibinfo{year}{2023}\natexlab{}.
\newblock \showarticletitle{A 28-nm 50.1-{TOPS/W} {P-8T} {SRAM}
  Compute-In-Memory Macro Design With {BL} Charge-Sharing-Based In-{SRAM}
  {DAC/ADC} Operations}.
\newblock \bibinfo{journal}{\emph{IEEE Journal of Solid-State Circuits (JSSC)}}
  (\bibinfo{year}{2023}).
\newblock


\bibitem[Li et~al\mbox{.}(2022)]%
        {li202240nm}
\bibfield{author}{\bibinfo{person}{Wantong Li}, \bibinfo{person}{James Read},
  \bibinfo{person}{Hongwu Jiang}, {and} \bibinfo{person}{Shimeng Yu}.}
  \bibinfo{year}{2022}\natexlab{}.
\newblock \showarticletitle{A 40nm {RRAM} Compute-in-Memory Macro with
  Parallelism-Preserving {ECC} for {Iso}-Accuracy Voltage Scaling}. In
  \bibinfo{booktitle}{\emph{IEEE European Solid State Circuits Conference
  (ESSCIRC)}}. IEEE, \bibinfo{pages}{101--104}.
\newblock


\bibitem[Liu et~al\mbox{.}(2020)]%
        {liu2020reactnet}
\bibfield{author}{\bibinfo{person}{Zechun Liu}, \bibinfo{person}{Zhiqiang
  Shen}, \bibinfo{person}{Marios Savvides}, {and} \bibinfo{person}{Kwang-Ting
  Cheng}.} \bibinfo{year}{2020}\natexlab{}.
\newblock \showarticletitle{{ReActNet}: Towards Precise Binary Neural Network
  with Generalized Activation Functions}. In \bibinfo{booktitle}{\emph{European
  Conference on Computer Vision (ECCV)}}.
\newblock


\bibitem[Nguyen et~al\mbox{.}(2021)]%
        {nguyen2021low}
\bibfield{author}{\bibinfo{person}{Thai-Hoang Nguyen},
  \bibinfo{person}{Muhammad Imran}, \bibinfo{person}{Jaehyuk Choi}, {and}
  \bibinfo{person}{Joon-Sung Yang}.} \bibinfo{year}{2021}\natexlab{}.
\newblock \showarticletitle{Low-cost and effective fault-tolerance enhancement
  techniques for emerging memories-based deep neural networks}. In
  \bibinfo{booktitle}{\emph{ACM/IEEE Design Automation Conference (DAC)}}.
  IEEE, \bibinfo{pages}{1075--1080}.
\newblock


\bibitem[Park et~al\mbox{.}(2014)]%
        {park2014fully}
\bibfield{author}{\bibinfo{person}{Youn~Sung Park}, \bibinfo{person}{Yaoyu
  Tao}, {and} \bibinfo{person}{Zhengya Zhang}.}
  \bibinfo{year}{2014}\natexlab{}.
\newblock \showarticletitle{A fully parallel nonbinary {LDPC} decoder with
  fine-grained dynamic clock gating}.
\newblock \bibinfo{journal}{\emph{IEEE Journal of Solid-State Circuits}}
  \bibinfo{volume}{50}, \bibinfo{number}{2} (\bibinfo{year}{2014}),
  \bibinfo{pages}{464--475}.
\newblock


\bibitem[Putra et~al\mbox{.}(2021)]%
        {putra2021sparkxd}
\bibfield{author}{\bibinfo{person}{Rachmad Vidya~Wicaksana Putra},
  \bibinfo{person}{Muhammad~Abdullah Hanif}, {and} \bibinfo{person}{Muhammad
  Shafique}.} \bibinfo{year}{2021}\natexlab{}.
\newblock \showarticletitle{{SparkXD}: A Framework for Resilient and
  Energy-Efficient Spiking Neural Network Inference using Approximate {DRAM}}.
  In \bibinfo{booktitle}{\emph{ACM/IEEE Design Automation Conference (DAC)}}.
  IEEE, \bibinfo{pages}{379--384}.
\newblock


\bibitem[Shin et~al\mbox{.}(2021)]%
        {shin2021fault}
\bibfield{author}{\bibinfo{person}{Hyein Shin}, \bibinfo{person}{Myeonggu
  Kang}, {and} \bibinfo{person}{Lee-Sup Kim}.} \bibinfo{year}{2021}\natexlab{}.
\newblock \showarticletitle{Fault-free: A Fault-resilient Deep Neural Network
  Accelerator based on Realistic {ReRAM} Devices}. In
  \bibinfo{booktitle}{\emph{ACM/IEEE Design Automation Conference (DAC)}}.
  IEEE, \bibinfo{pages}{1039--1044}.
\newblock


\bibitem[Tao and Wu(2019)]%
        {2019automated}
\bibfield{author}{\bibinfo{person}{Yaoyu Tao} {and} \bibinfo{person}{Qi Wu}.}
  \bibinfo{year}{2019}\natexlab{}.
\newblock \showarticletitle{An Automated {FPGA}-Based Framework for Rapid
  Prototyping of Nonbinary {LDPC} Codes}. In \bibinfo{booktitle}{\emph{IEEE
  International Symposium on Circuits and Systems (ISCAS)}}. IEEE,
  \bibinfo{pages}{1--5}.
\newblock


\bibitem[Venkiah et~al\mbox{.}(2008)]%
        {venkiah2008design}
\bibfield{author}{\bibinfo{person}{Auguste Venkiah}, \bibinfo{person}{David
  Declercq}, {and} \bibinfo{person}{Charly Poulliat}.}
  \bibinfo{year}{2008}\natexlab{}.
\newblock \showarticletitle{Design of cages with a randomized progressive
  edge-growth algorithm}.
\newblock \bibinfo{journal}{\emph{IEEE Communications Letters}}
  \bibinfo{volume}{12}, \bibinfo{number}{4} (\bibinfo{year}{2008}),
  \bibinfo{pages}{301--303}.
\newblock


\bibitem[Wen et~al\mbox{.}(2024)]%
        {wen2024fusion}
\bibfield{author}{\bibinfo{person}{Tai-Hao Wen}, \bibinfo{person}{Je-Min Hung},
  \bibinfo{person}{Wei-Hsing Huang}, \bibinfo{person}{Chuan-Jia Jhang},
  \bibinfo{person}{Yun-Chen Lo}, \bibinfo{person}{Hung-Hsi Hsu},
  \bibinfo{person}{Zhao-En Ke}, \bibinfo{person}{Yu-Chiao Chen},
  \bibinfo{person}{Yu-Hsiang Chin}, \bibinfo{person}{Chin-I Su},
  {et~al\mbox{.}}} \bibinfo{year}{2024}\natexlab{}.
\newblock \showarticletitle{Fusion of memristor and digital compute-in-memory
  processing for energy-efficient edge computing}.
\newblock \bibinfo{journal}{\emph{Science}} \bibinfo{volume}{384},
  \bibinfo{number}{6693} (\bibinfo{year}{2024}), \bibinfo{pages}{325--332}.
\newblock


\bibitem[Wu et~al\mbox{.}(2018)]%
        {jing2022vsdca}
\bibfield{author}{\bibinfo{person}{Tony~F Wu}, \bibinfo{person}{Haitong Li},
  \bibinfo{person}{Ping-Chen Huang}, \bibinfo{person}{Abbas Rahimi},
  \bibinfo{person}{Gage Hills}, \bibinfo{person}{Bryce Hodson},
  \bibinfo{person}{William Hwang}, \bibinfo{person}{Jan~M Rabaey},
  \bibinfo{person}{H-S~Philip Wong}, \bibinfo{person}{Max~M Shulaker},
  {et~al\mbox{.}}} \bibinfo{year}{2018}\natexlab{}.
\newblock \showarticletitle{Hyperdimensional computing exploiting carbon
  nanotube {FET}s, resistive {RAM}, and their monolithic {3D} integration}.
\newblock \bibinfo{journal}{\emph{IEEE Journal of Solid-State Circuits}}
  \bibinfo{volume}{53}, \bibinfo{number}{11} (\bibinfo{year}{2018}),
  \bibinfo{pages}{3183--3196}.
\newblock


\bibitem[Xie et~al\mbox{.}(2024)]%
        {xie2024ed}
\bibfield{author}{\bibinfo{person}{Wenao Xie}, \bibinfo{person}{Haoyang Sang},
  \bibinfo{person}{Beomseok Kwon}, \bibinfo{person}{Dongseok Im},
  \bibinfo{person}{Sangjin Kim}, \bibinfo{person}{Sangyeob Kim},
  \bibinfo{person}{Kangho Lee}, {and} \bibinfo{person}{Hoi-Jun Yoo}.}
  \bibinfo{year}{2024}\natexlab{}.
\newblock \showarticletitle{{ED-MPIM}: An Energy-Efficient Event-Driven Smart
  Vision {SoC} With High-Linearity and Reconfigurable {MRAM} PIM}.
\newblock \bibinfo{journal}{\emph{IEEE Journal of Solid-State Circuits (JSSC)}}
  (\bibinfo{year}{2024}).
\newblock


\bibitem[Yan et~al\mbox{.}(2022)]%
        {9731545}
\bibfield{author}{\bibinfo{person}{Bonan Yan}, \bibinfo{person}{Jeng-Long Hsu},
  \bibinfo{person}{Pang-Cheng Yu}, \bibinfo{person}{Chia-Chi Lee},
  \bibinfo{person}{Yaojun Zhang}, \bibinfo{person}{Wenshuo Yue},
  \bibinfo{person}{Guoqiang Mei}, \bibinfo{person}{Yuchao Yang},
  \bibinfo{person}{Yue Yang}, \bibinfo{person}{Hai Li}, \bibinfo{person}{Yiran
  Chen}, {and} \bibinfo{person}{Ru Huang}.} \bibinfo{year}{2022}\natexlab{}.
\newblock \showarticletitle{A 1.041-{M}b/mm2 27.38-{TOPS/W} Signed-{INT8}
  Dynamic-Logic-Based {ADC}-less {SRAM} Compute-in-Memory Macro in 28nm with
  Reconfigurable Bitwise Operation for {AI} and Embedded Applications}. In
  \bibinfo{booktitle}{\emph{2022 IEEE International Solid-State Circuits
  Conference (ISSCC)}}, Vol.~\bibinfo{volume}{65}. \bibinfo{pages}{188--190}.
\newblock
\href{https://doi.org/10.1109/ISSCC42614.2022.9731545}{doi:\nolinkurl{10.1109/ISSCC42614.2022.9731545}}


\bibitem[Yan et~al\mbox{.}(2019a)]%
        {aisy.201900068}
\bibfield{author}{\bibinfo{person}{Bonan Yan}, \bibinfo{person}{Bing Li},
  \bibinfo{person}{Ximing Qiao}, \bibinfo{person}{Cheng-Xin Xue},
  \bibinfo{person}{Meng-Fan Chang}, \bibinfo{person}{Yiran Chen}, {and}
  \bibinfo{person}{Hai~(Helen) Li}.} \bibinfo{year}{2019}\natexlab{a}.
\newblock \showarticletitle{Resistive Memory-Based In-Memory Computing: From
  Device and Large-Scale Integration System Perspectives}.
\newblock \bibinfo{journal}{\emph{Advanced Intelligent Systems}}
  \bibinfo{volume}{1}, \bibinfo{number}{7} (\bibinfo{year}{2019}),
  \bibinfo{pages}{1900068}.
\newblock
\href{https://doi.org/10.1002/aisy.201900068}{doi:\nolinkurl{10.1002/aisy.201900068}}
\showeprint{https://advanced.onlinelibrary.wiley.com/doi/pdf/10.1002/aisy.201900068}


\bibitem[Yan et~al\mbox{.}(2019b)]%
        {8776485}
\bibfield{author}{\bibinfo{person}{Bonan Yan}, \bibinfo{person}{Qing Yang},
  \bibinfo{person}{Wei-Hao Chen}, \bibinfo{person}{Kung-Tang Chang},
  \bibinfo{person}{Jian-Wei Su}, \bibinfo{person}{Chien-Hua Hsu},
  \bibinfo{person}{Sih-Han Li}, \bibinfo{person}{Heng-Yuan Lee},
  \bibinfo{person}{Shyh-Shyuan Sheu}, \bibinfo{person}{Mon-Shu Ho},
  \bibinfo{person}{Qing Wu}, \bibinfo{person}{Meng-Fan Chang},
  \bibinfo{person}{Yiran Chen}, {and} \bibinfo{person}{Hai Li}.}
  \bibinfo{year}{2019}\natexlab{b}.
\newblock \showarticletitle{{RRAM}-based Spiking Nonvolatile
  Computing-In-Memory Processing Engine with Precision-Configurable In Situ
  Nonlinear Activation}. In \bibinfo{booktitle}{\emph{2019 Symposium on VLSI
  Technology}}. \bibinfo{pages}{T86--T87}.
\newblock
\href{https://doi.org/10.23919/VLSIT.2019.8776485}{doi:\nolinkurl{10.23919/VLSIT.2019.8776485}}


\bibitem[Yang et~al\mbox{.}(2020)]%
        {yang2020retransformer}
\bibfield{author}{\bibinfo{person}{Xiaoxuan Yang}, \bibinfo{person}{Bonan Yan},
  \bibinfo{person}{Hai Li}, {and} \bibinfo{person}{Yiran Chen}.}
  \bibinfo{year}{2020}\natexlab{}.
\newblock \showarticletitle{{ReTransformer}: {ReRAM}-based processing-in-memory
  architecture for transformer acceleration}. In
  \bibinfo{booktitle}{\emph{International Conference on Computer-Aided Design
  (ICCAD)}}. \bibinfo{pages}{1--9}.
\newblock


\bibitem[Yin et~al\mbox{.}(2024)]%
        {zheng2024improving}
\bibfield{author}{\bibinfo{person}{Xunzhao Yin}, \bibinfo{person}{Franz
  M{\"u}ller}, \bibinfo{person}{Ann~Franchesca Laguna}, \bibinfo{person}{Chao
  Li}, \bibinfo{person}{Qingrong Huang}, \bibinfo{person}{Zhiguo Shi},
  \bibinfo{person}{Maximilian Lederer}, \bibinfo{person}{Nellie Laleni},
  \bibinfo{person}{Shan Deng}, \bibinfo{person}{Zijian Zhao}, {et~al\mbox{.}}}
  \bibinfo{year}{2024}\natexlab{}.
\newblock \showarticletitle{Deep random forest with ferroelectric analog
  content addressable memory}.
\newblock \bibinfo{journal}{\emph{Science Advances}} \bibinfo{volume}{10},
  \bibinfo{number}{23} (\bibinfo{year}{2024}), \bibinfo{pages}{eadk8471}.
\newblock


\bibitem[Yoon et~al\mbox{.}(2022)]%
        {yoon202240}
\bibfield{author}{\bibinfo{person}{Jong-Hyeok Yoon}, \bibinfo{person}{Muya
  Chang}, \bibinfo{person}{Win-San Khwa}, \bibinfo{person}{Yu-Der Chih},
  \bibinfo{person}{Meng-Fan Chang}, {and} \bibinfo{person}{Arijit
  Raychowdhury}.} \bibinfo{year}{2022}\natexlab{}.
\newblock \showarticletitle{A 40-nm 118.44-{TOPS/W} Voltage-Sensing
  Compute-in-Memory {RRAM} Macro With Write Verification and Multi-Bit
  Encoding}.
\newblock \bibinfo{journal}{\emph{IEEE Journal of Solid-State Circuits (JSSC)}}
  \bibinfo{volume}{57}, \bibinfo{number}{3} (\bibinfo{year}{2022}),
  \bibinfo{pages}{845--857}.
\newblock


\bibitem[Yue et~al\mbox{.}(2025)]%
        {yue2025physical}
\bibfield{author}{\bibinfo{person}{Wenshuo Yue}, \bibinfo{person}{Kai Wu},
  \bibinfo{person}{Zhiyuan Li}, \bibinfo{person}{Juchen Zhou},
  \bibinfo{person}{Zeyu Wang}, \bibinfo{person}{Teng Zhang},
  \bibinfo{person}{Yuxiang Yang}, \bibinfo{person}{Lintao Ye},
  \bibinfo{person}{Yongqin Wu}, \bibinfo{person}{Weihai Bu}, {et~al\mbox{.}}}
  \bibinfo{year}{2025}\natexlab{}.
\newblock \showarticletitle{Physical unclonable in-memory computing for
  simultaneous protecting private data and deep learning models}.
\newblock \bibinfo{journal}{\emph{Nature Communications}} \bibinfo{volume}{16},
  \bibinfo{number}{1} (\bibinfo{year}{2025}), \bibinfo{pages}{1031}.
\newblock


\bibitem[Yue et~al\mbox{.}(2024)]%
        {yue2024scalable}
\bibfield{author}{\bibinfo{person}{Wenshuo Yue}, \bibinfo{person}{Teng Zhang},
  \bibinfo{person}{Zhaokun Jing}, \bibinfo{person}{Kai Wu},
  \bibinfo{person}{Yuxiang Yang}, \bibinfo{person}{Zhen Yang},
  \bibinfo{person}{Yongqin Wu}, \bibinfo{person}{Weihai Bu},
  \bibinfo{person}{Kai Zheng}, \bibinfo{person}{Jin Kang}, {et~al\mbox{.}}}
  \bibinfo{year}{2024}\natexlab{}.
\newblock \showarticletitle{A scalable universal {Ising} machine based on
  interaction-centric storage and compute-in-memory}.
\newblock \bibinfo{journal}{\emph{Nature Electronics}} \bibinfo{volume}{7},
  \bibinfo{number}{10} (\bibinfo{year}{2024}), \bibinfo{pages}{904--913}.
\newblock


\bibitem[Zhang et~al\mbox{.}(2024)]%
        {zhang2024efficient}
\bibfield{author}{\bibinfo{person}{Fan Zhang}, \bibinfo{person}{Amitesh
  Sridharan}, \bibinfo{person}{Wilman Tsai}, \bibinfo{person}{Yiran Chen},
  \bibinfo{person}{Shan~X Wang}, {and} \bibinfo{person}{Deliang Fan}.}
  \bibinfo{year}{2024}\natexlab{}.
\newblock \showarticletitle{Efficient Memory Integration: {MRAM-SRAM} Hybrid
  Accelerator for Sparse On-Device Learning}. In
  \bibinfo{booktitle}{\emph{ACM/IEEE Design Automation Conference (DAC)}}.
  \bibinfo{pages}{1--6}.
\newblock


\bibitem[Zhang et~al\mbox{.}(2021)]%
        {zhang2021efficient}
\bibfield{author}{\bibinfo{person}{Grace~Li Zhang}, \bibinfo{person}{Bing Li},
  \bibinfo{person}{Xing Huang}, \bibinfo{person}{Chen Shen},
  \bibinfo{person}{Shuhang Zhang}, \bibinfo{person}{Florin Burcea},
  \bibinfo{person}{Helmut Graeb}, \bibinfo{person}{Tsung-Yi Ho},
  \bibinfo{person}{Hai Li}, {and} \bibinfo{person}{Ulf Schlichtmann}.}
  \bibinfo{year}{2021}\natexlab{}.
\newblock \showarticletitle{An efficient programming framework for
  memristor-based neuromorphic computing}. In \bibinfo{booktitle}{\emph{Design,
  Automation \& Test in Europe Conference \& Exhibition (DATE)}}. IEEE,
  \bibinfo{pages}{1068--1073}.
\newblock


\bibitem[Zhang et~al\mbox{.}(2019)]%
        {zhang2019design}
\bibfield{author}{\bibinfo{person}{Wenqiang Zhang}, \bibinfo{person}{Xiaochen
  Peng}, \bibinfo{person}{Huaqiang Wu}, \bibinfo{person}{Bin Gao},
  \bibinfo{person}{Hu He}, \bibinfo{person}{Youhui Zhang},
  \bibinfo{person}{Shimeng Yu}, {and} \bibinfo{person}{He Qian}.}
  \bibinfo{year}{2019}\natexlab{}.
\newblock \showarticletitle{Design guidelines of {RRAM} based
  neural-processing-unit: A joint device-circuit-algorithm analysis}. In
  \bibinfo{booktitle}{\emph{ACM/IEEE Design Automation Conference (DAC)}}.
  \bibinfo{pages}{1--6}.
\newblock


\bibitem[Zhou et~al\mbox{.}(2016)]%
        {zhou2016dorefa}
\bibfield{author}{\bibinfo{person}{Shuchang Zhou}, \bibinfo{person}{Yuxin Wu},
  \bibinfo{person}{Zekun Ni}, \bibinfo{person}{Xinyu Zhou}, \bibinfo{person}{He
  Wen}, {and} \bibinfo{person}{Yuheng Zou}.} \bibinfo{year}{2016}\natexlab{}.
\newblock \showarticletitle{Dorefa-net: Training low bitwidth convolutional
  neural networks with low bitwidth gradients}.
\newblock \bibinfo{journal}{\emph{arXiv preprint arXiv:1606.06160}}
  (\bibinfo{year}{2016}).
\newblock


\end{thebibliography}

\end{document}